\documentstyle[sprocl,epsfig]{article}

\newcommand {\ignore}[1]{}

\newcommand{\bc}{\begin{center}}
\newcommand{\ec}{\end{center}}
\newcommand{\epm}{e^+e^-}
\def\ifmath#1{\relax\ifmmode #1\else $#1$\fi}
\def\VEV#1{\left\langle #1\right\rangle}
\def\mathrm#1{\hbox{#1}}
\def\ra{\rightarrow}

\def\nt{\hbox{$\nu_\tau$ }}

\def\mnt{\hbox{$m_{\nu_\tau}$ }}

\def\eq#1{{eq. (\ref{#1})}}

\def\lsim{\raise0.3ex\hbox{$\;<$\kern-0.75em\raise-1.1ex\hbox{$\sim\;$}}}
\def\gsim{\raise0.3ex\hbox{$\;>$\kern-0.75em\raise-1.1ex\hbox{$\sim\;$}}}
\def\bel{\begin{letter}}
\def\eel{\end{letter}}
\def\beq{\begin{equation}}
\def\eeq{\end{equation}}
\def\bef{\begin{figure}}
\def\eef{\end{figure}}
\def\bet{\begin{table}}
\def\eet{\end{table}}
\def\bea{\begin{eqnarray}}
\def\ba{\begin{array}}
\def\ea{\end{array}}
\def\bi{\begin{itemize}}
\def\ei{\end{itemize}}
\def\ben{\begin{enumerate}}
\def\een{\end{enumerate}}
\def\ra{\rightarrow}

\def\eea{\end{eqnarray}}
\def\ib#1#2#3{           {\it ibid. }{\bf #1} (19#2) #3}

\def\nps#1#2#3{          {\it Nucl. Phys. B (Proc. Suppl.) }
                         {\bf #1} (19#2) #3}
\def\np#1#2#3{           {\it Nucl. Phys. }{\bf #1} (19#2) #3}
\def\pl#1#2#3{           {\it Phys. Lett. }{\bf #1} (19#2) #3}
\def\pr#1#2#3{           {\it Phys. Rev. }{\bf #1} (19#2) #3}
\def\prep#1#2#3{         {\it Phys. Rep. }{\bf #1} (19#2) #3}
\def\prl#1#2#3{          {\it Phys. Rev. Lett. }{\bf #1} (19#2) #3}

\def\zp#1#2#3{           {\it Zeit. f\"ur Physik }{\bf #1} (19#2) #3}

\def\ppnp#1#2#3{           {\it Prog. Part. Nucl. Phys. }{\bf #1} (19#2) #3}

\relax

\begin{document}

\title{EFFECTS OF R PARITY BREAKING IN THE HIGGS SECTOR}

\author{J. ROSIEK}

\address{Institut f\"ur Theoretische Physik, Universit\"at Karlsruhe}

\maketitle\abstracts{
We discuss possible manifestations of R parity breaking in the Higgs
sector. We illustrate this with three examples: high-energy mono-photon
production at LEP,
calculation of limits on
associated production of invisibly decaying Higgs bosons
and discussion of
non-standard scalar decays in the MSSM with broken R parity.}

\baselineskip=15.9pt

\vspace{-8.5cm}
\begin{flushright}
\begin{minipage}[t]{3cm}
KA-TP-7-1995\\
hep-ph/9508355\\
August 1995
\end{minipage}
\end{flushright}
\vspace{6cm}

\section{Introduction}

In analysing the supersymmetric theories
one usually assumes the conservation of a discrete symmetry, called R
parity\cite{wein}, distinguishing matter fields from their superpartners.
The R parity violation could lead to phenomenologically interesting
effects, such as mixing of gauginos with leptons\cite{valleross} and Higgs
bosons with sleptons\cite{mas1,asj2}
(and hence breaking of the lepton number), non-zero
neutrino masses\cite{hall,asj2} and their decays\cite{RPMSW},
$Z^0$ decays to single charginos or neutralinos\cite{ROMA},
existence of a massless Goldstone boson called majoron\cite{CMP}
and many others.
In contrast to the R parity breaking in the fermion sector,
its consequences in the scalar sector are relatively less
explored. We present three examples of possible manifestations
of R-parity breaking in the Higgs sector:
high-energy mono-photon production at LEP\cite{MY3},
associated production of invisibly decaying Higgs bosons\cite{MY1}
and novel scalar decays in the MSSM with broken R parity\cite{MY2}.

\section{Single Photon Decays of the $Z^0$ and SUSY with Broken R-Parity}

Recently the OPAL collaboration
has published a high statistics single photon spectrum that shows
some excess of high energy photons above the expectations from g
the initial state radiation (ISR)\cite{opal}.
This excess could be explained in a class of SUSY
models with spontaneous violation of R parity,
\begin{figure}[htbp]
\begin{center}
\vspace*{-3mm}
\mbox{
\epsfig{figure=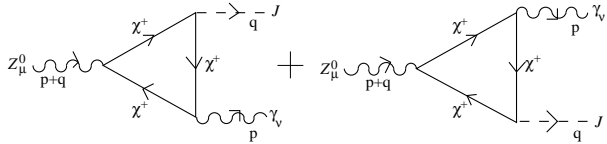,height=0.16\linewidth}
}
\end{center}
\vspace*{-5mm}
\caption{Diagrams contributing to the monochromatic
 single photon emission $Z^0 \rightarrow \gamma J$
\label{fig:zfj1}}
\vspace*{-4mm}
\end{figure}
which predict the existence of new $Z^0$ decay modes
involving single ($Z^0 \rightarrow J \gamma$) or double
($Z^0 \rightarrow JJ \gamma$) invisible massless majoron emission.

Single majoron emission would produce
monochromatic photons emitted with energy $\frac{M_Z}{2}$.
In order to estimate the attainable values of Br($Z^0 \rightarrow J \gamma$)
we adopt as a model for the spontaneous R parity violation
the one proposed in\cite{MASI_pot3}. Diagrams contributing at the 1-loop
level to the process $Z^0 \rightarrow J \gamma$ are shown in
Fig.~\ref{fig:zfj1}. We obtained the explicit expression for
Br($Z^0\rightarrow\gamma J$) by
calculation of the 3-point Green functions of Fig.~\ref{fig:zfj1}.
We have varied model pa\-ra\-me\-ters, rejecting points
violating constraints imposed by existing
laboratory observations, cosmology and astrophysics.
We have found that the branching ratio for the 2-body $Z^0$ decay
to a single photon + missing momentum in the considered
model can reach $10^{-5}$, being within the sensitivity of LEP\cite{MY3}.

Double emission process $Z^0 \rightarrow JJ \gamma$ would give rise to a
continuous photon spectrum\footnote{
We do not consider here other possible mechanisms, like
LSP pair production followed by the radiative
decay of one of them $\chi^0 \ra \gamma \nu$ and the invisible decay
of the other $\chi^0 \ra J \nu$.}.
Imposing gauge and $CP$ invariance and Bose symmetry,
one can express the on-shell amplitude
in terms of a single form-factor $V_{0}$:
\begin{equation}
A (Z^0 \rightarrow JJ \gamma ) =  V_{0}
[p^{\mu} (q_{1} + q_{2})^{\nu} - p (q_{1} + q_{2})g^{\mu \nu} ]
\epsilon_{\mu} (p+q_1+q_2) \epsilon_{\nu} (p)
\label{amplitude}
\end{equation}
As the simplest illustration we derive the
$\gamma$-spectrum following from eq.~(\ref{amplitude}) in the approximation of
constant $V_{0}$. We obtain
\begin{equation}
\frac{d \Gamma}{d E_{\gamma}} = \frac{|V_{0}|^{2} M_Z}{96 \pi^{3}}
E_{\gamma}^{3}~~~~~~~~~~~~0 \leq E_{\gamma} \leq \frac{M_Z}{2}
\end{equation}
The details of the shape of the $\gamma$ spectrum depend upon the specific
model and SUSY parameters choice, but its general characteristic is
determined by kinematics and is
different from the SM process $Z^0 \rightarrow \nu \overline{\nu} \gamma$.
Superimposed upon the continuous spectrum we have, in addition,
a spike at its endpoint, corresponding to the emission of the
monochromatic $\gamma $ in the $Z^0 \rightarrow \gamma J $ decay.

\section{Limits on Associated Production of
Invisibly Decaying Higgs Bosons from $Z^0$ Decays}

In a class of models with a spontaneously broken global
symmetry the CP-even Higgs boson(s) are expected to have sizeable invisible
decay modes to the majorons\cite{bj}. This decay
could contribute to the signal looked for at LEP -- two
acoplanar jets + missing momentum.
The existing analyses of the invisible Higgs search
have concentrated on a minimally extended SM Higgs sector with the
addition of a Higgs singlet and hence on the Bjorken process\cite{alfonso}.
We extend this analysis to include the model containing two
Higgs doublets and a singlet (carrying lepton number)\cite{MASI_pot3}
and the associated production of the CP-even and CP-odd Higgs
bosons\cite{MY1}.
For the illustration, we display the constraints
obtained for the published ALEPH data sample\cite{Aleph92}
based on a statistics of $1.23\times 10^6$  hadronic $Z^0$ events.

After spontaneous $SU(2) \times U(1) \times U(1)_L$
breaking, the Higgs sector of the considered model contains
3 massive CP-even scalars $H_i$,
a massive CP-odd scalar $A$  and
the massless majoron $J$. We assume that at LEP only the
lightest CP-even scalar $H_1 \equiv H$ can be produced.
The relevant couplings are:
\begin{eqnarray}
\label{HZZ3}
{\cal L}_{HZZ}&=&(\sqrt{2}G_F)^{1/2}M_Z^2\;\epsilon_B \;Z_{\mu}Z^{\mu} H\\
{\cal L}_{HAZ}&=&-{e\over \sin\theta_W\cos\theta_W}\;\epsilon_A\; Z_{\mu}H
\stackrel{\leftrightarrow}{\partial^\mu}A.
\end{eqnarray}
where $\epsilon_A^2,\epsilon_B^2\leq 1$.
The main $H$ decay modes are $b\overline b$, $JJ$ and $AA$
(if $m_H>2m_A$).
$HJJ$ coupling strength is unconstrained
and can be effectively parameterized by
$B$ = Br$(H \ra JJ)$, $0\leq B \leq 1$\footnote{The parameterization of the
expected number of the invisible Higgs boson events in terms of
$m_H$, $m_A$, $\epsilon_A$, $\epsilon_B$ and $B$ is quite general
and not limited only to the model\cite{MASI_pot3}.}.
The decay $A\ra JJJ$ does not exist since the (CP-allowed) couplings
$AJ^3$ or $AHJ$ vanish
at the tree level in our model.
\begin{figure}[htbp]
\vspace*{-7mm}
\begin{tabular}{p{0.468\linewidth}p{0.468\linewidth}}
\begin{center}
\mbox{
\epsfig{file=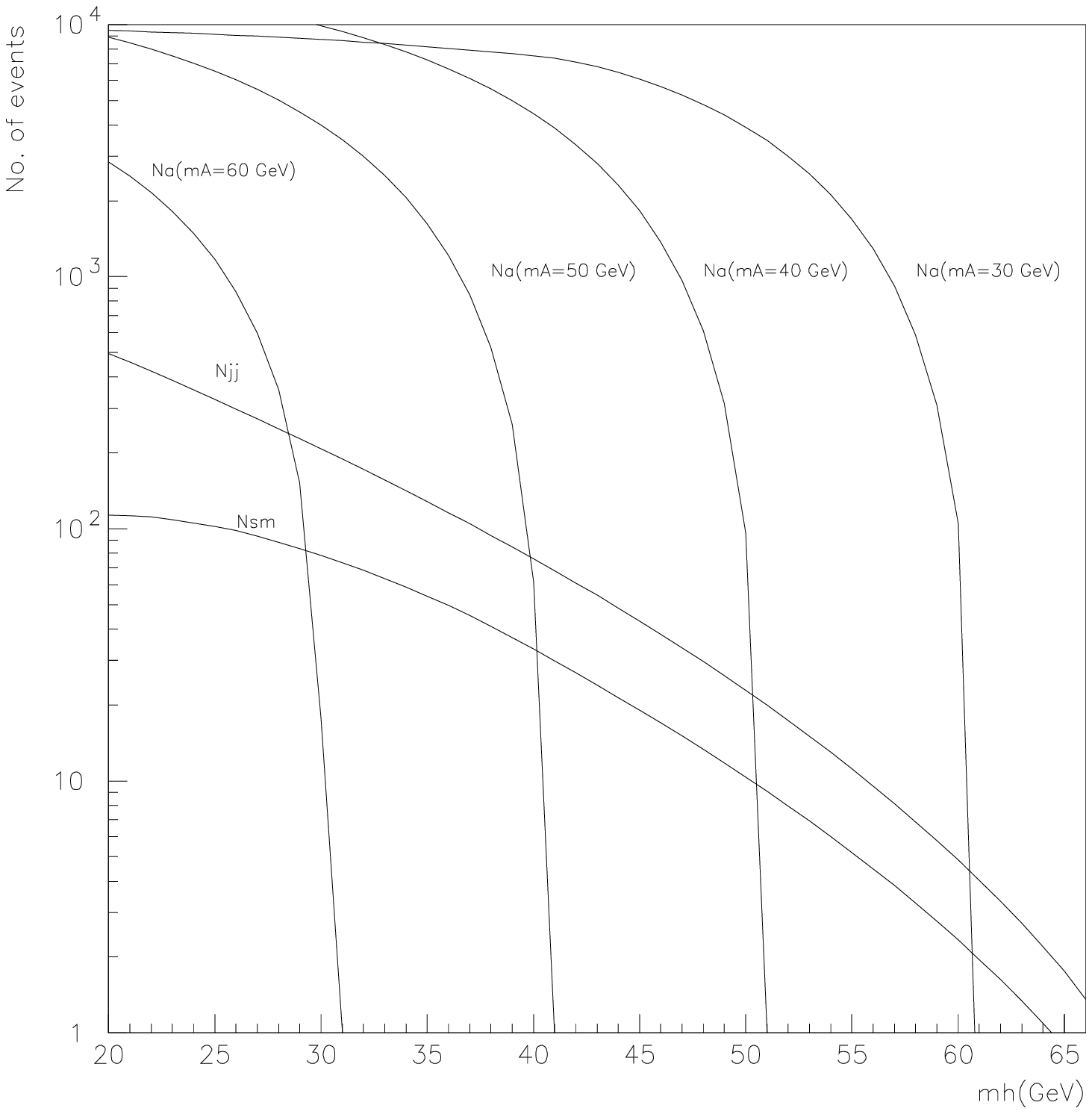,width=0.95\linewidth,height=0.9\linewidth}}
\end{center}  &
\begin{center}
\mbox{
\epsfig{file=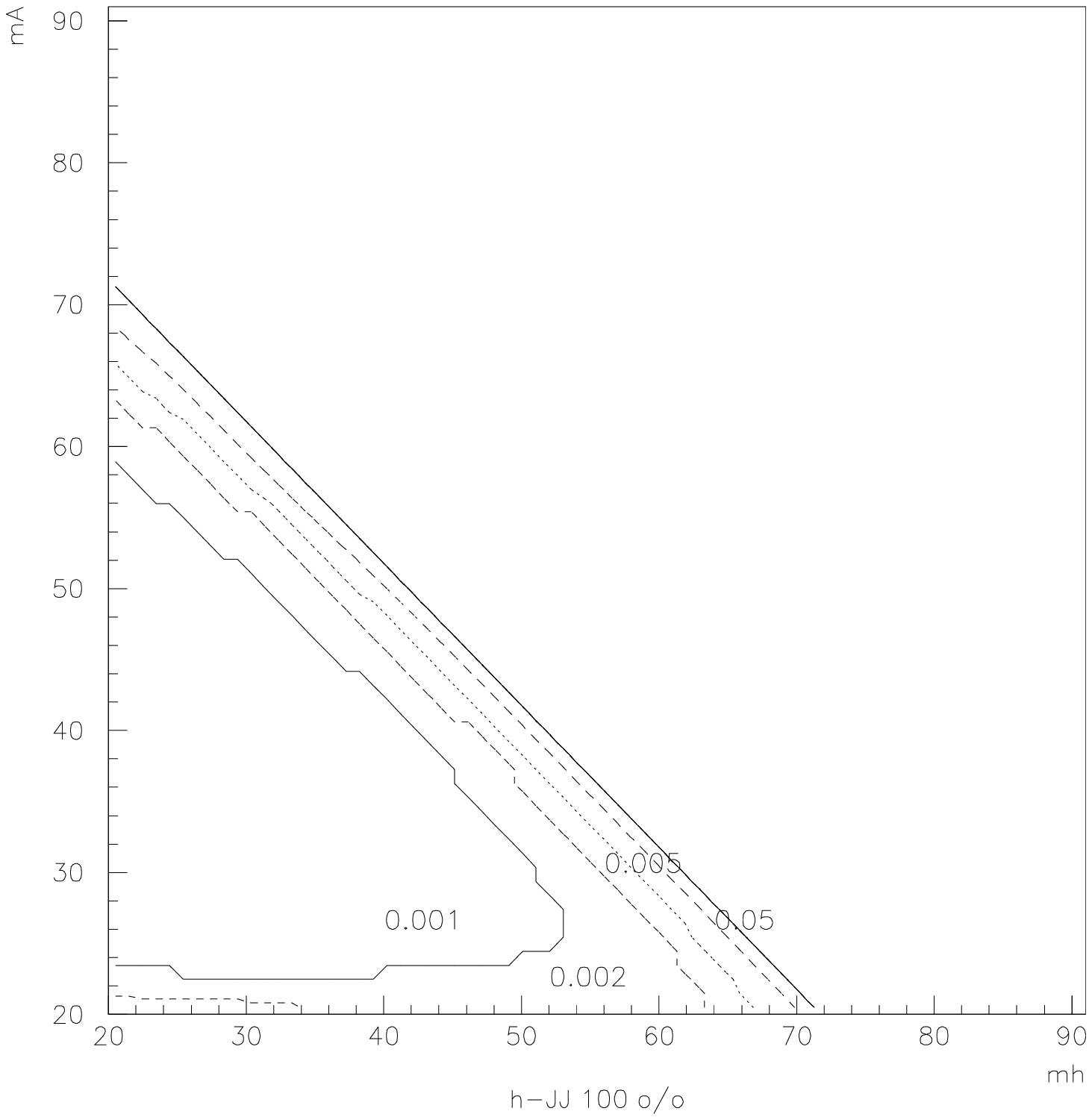,width=0.95\linewidth,height=0.9\linewidth}}
\end{center} \\
\vspace*{-12mm}
\caption{
Numbers of expected dijet + missing momentum events
after imposing ALEPH cuts.
\label{fig:dp1}
}&
\vspace*{-12mm}
\caption{
Limits on $\epsilon^2_{A}$ in the $m_A$-$m_H$ plane based on
$e^+ e^- \rightarrow HA \rightarrow JJ b\overline b$ channel.
We have assumed Br($H \rightarrow JJ$) = 100\%.
\label{fig:dp3}
} \\
\end{tabular}
\vspace*{-12mm}
\end{figure}
Since $A$ can decay only visibly ($B_A=$ Br$(A\rightarrow b \overline{b})
\approx 90\%$ for $m_A \geq 20$ GeV), one expects dijets +
missing momentum as a signal of the Higgs production.
This signal arises from three processes:
1) $Z^{0\star} \rightarrow  \nu \overline{\nu}$,
$H \rightarrow b \overline{b}$,
2) $Z^{0\star} \rightarrow q \overline{q}$, $H \ra JJ$
and 3) $H\rightarrow JJ$, $A\rightarrow b \overline{b}$.
For each process one has a sizeable missing momentum which is aligned
neither along the beam nor along the jets.  For the SM
background missing momentum arises from i) jet fluctuation
(including escaping $\nu$ from $b,c$ and $\tau$ jets) in which case it
is aligned along the jets, and ii) ISR or
$e^+e^- \rightarrow (e^+e^-) \gamma\gamma$ process in
which case it is aligned along the beam direction. This enables one to
eliminate the SM background by a combination of kinematic cuts\cite{aleph1}.
We denote the number of expected signal events for the processes 1)-3), after
the cuts, by $N_{SM}$, $N_{JJ}$ and $N_A$ respectively, assuming no
suppression due to the mixing angles or branching fractions in each
case.  The expected number of signal events, after incorporating
these effects, is given by
\begin{equation}
N_{2j}=\epsilon_B^2\left[ B N_{J}+(1-B) N_{SM}\right]+\epsilon_A^2
B_A B N_A
\end{equation}
As can be seen in Fig.~\ref{fig:dp1}, $N_A \gg N_{SM}, N_{JJ}$,
implying that, if not suppressed kinematically or by mixing angles,
associated production gives the strong limits
on $\epsilon_A^2$ (see Fig.~\ref{fig:dp3},
where we have assumed fully invisible $H$ decay).
The limit on $\epsilon_B^2$ is given by the Bjorken process and is
the same as obtained in\cite{alfonso}.
A decay mode independent limits on $\epsilon_A^2$,
$\epsilon_B^2$ can be obtained by varying $B$
from 0 to 1 and combining
the data on dijet + missing momentum with those from 4
and 6 $b$-jet searches. The overall limit is
dominated by visible channels, which give weaker constraints
on $\epsilon_A^2$, $\epsilon_B^2$ then those obtained for the invisible
$H$ decay.

\section{Novel Scalar Boson Decays in SUSY with Broken R-Parity}

The standard MSSM R parity conserving superpotential can be generalized
by adding the following R parity (and lepton number) violating terms:
\beq \label{wr}
W_R=\epsilon_{ab}\left[\lambda_{ijk}\hat{L}_i^a \hat{L}_j^b
\hat{E}_k^C + \lambda'_{ijk}\hat{L}_i^a \hat{Q}_j^b \hat{D}_k^C
+ \epsilon_i \hat{L}_i^a \hat{H}_2^b \right]
\end{equation}
We focus on the effect of the last term in \eq{wr},
assuming in addition $\epsilon_{1,2}\approx 0$.
%, as they are
%strongly constrained by the limits on \ne and \nm masses.
The remaining $\epsilon_3\hat{L}_3 \hat{H}_2$
term induces a non-zero VEV for the tau sneutrino:
%\footnote{
%The $\hat{L}_3$ and $\hat{H}_1$ superfields can be redefined
%so as to eliminate the $\hat{L}_3 \hat{H}_2$ term
%from the superpotential. However,
%such redefinition generates a bilinear term in the scalar soft SUSY-breaking
%potential\cite{MY2}.}:
$\VEV{\tilde{\nu_{\tau}}} \equiv \frac{v_3}{\sqrt 2}$ and
leads to mixing of gauginos with leptons, Higgs scalars
with the tau sneutrino and hence
to the R parity violating Higgs bosons decay
modes.

All the elements of the various mixing matrices can be expressed
in terms of six independent parameters
which we choose as
$\tan \beta = \frac{v_2}{\sqrt{v_1^2 + v_3^2}}$,
$\mu$, $\epsilon_3$, $m_A^2$, the gaugino and soft sneutrino
mass parameters $M_2$ and $m_{L_3}$.
We have taken into account the following constraints on the model parameters:

\noindent 1) The major constraint on
$\epsilon_3$ and hence on R parity violating mixings comes
from bound on \nt mass\cite{PDG94}, induced by the non-zero
$\epsilon_3$: $\mnt\leq 30$ MeV\footnote{see
comments in\cite{MY2} on possible cosmological constraints.}.

\noindent 2) Decay widths $\Gamma(Z^0\ra \chi^0\chi^0,\chi^+\chi^-)$
should obey the LEP restrictions.

\noindent The decay $h\ra \chi^0 \nu_\tau$ occurs either through the
sneutrino component of $h$,
\beq
h=a_{31} (\tilde{\nu}_\tau)_R + a_{21} (\phi_2)_R + a_{11} (\phi_1)_R
\label{a31}
\eeq
or through $\nu_{\tau}$ admixture in the LSP in the
$h \chi\chi$ vertex. The mixing $a_{31}$ appearing in~\eq{a31}
is of the order of
${\cal O}\left(\frac{\mu\epsilon_3}{m_h^2 -
m_{\tilde{\nu}_\tau}^2}\right)$ and
become large if the {\sl sneutrino} mass is close to the mass of the
relevant {\sl Higgs} boson.
The relative importance of the SUSY decay mode
$h \ra \chi^0 \nu_\tau$ follows from the ratio:
\begin{figure}[htbp]
\vspace*{-8mm}
\begin{tabular}{p{0.468\linewidth}p{0.468\linewidth}}
\begin{center}
\mbox{
\epsfig{file=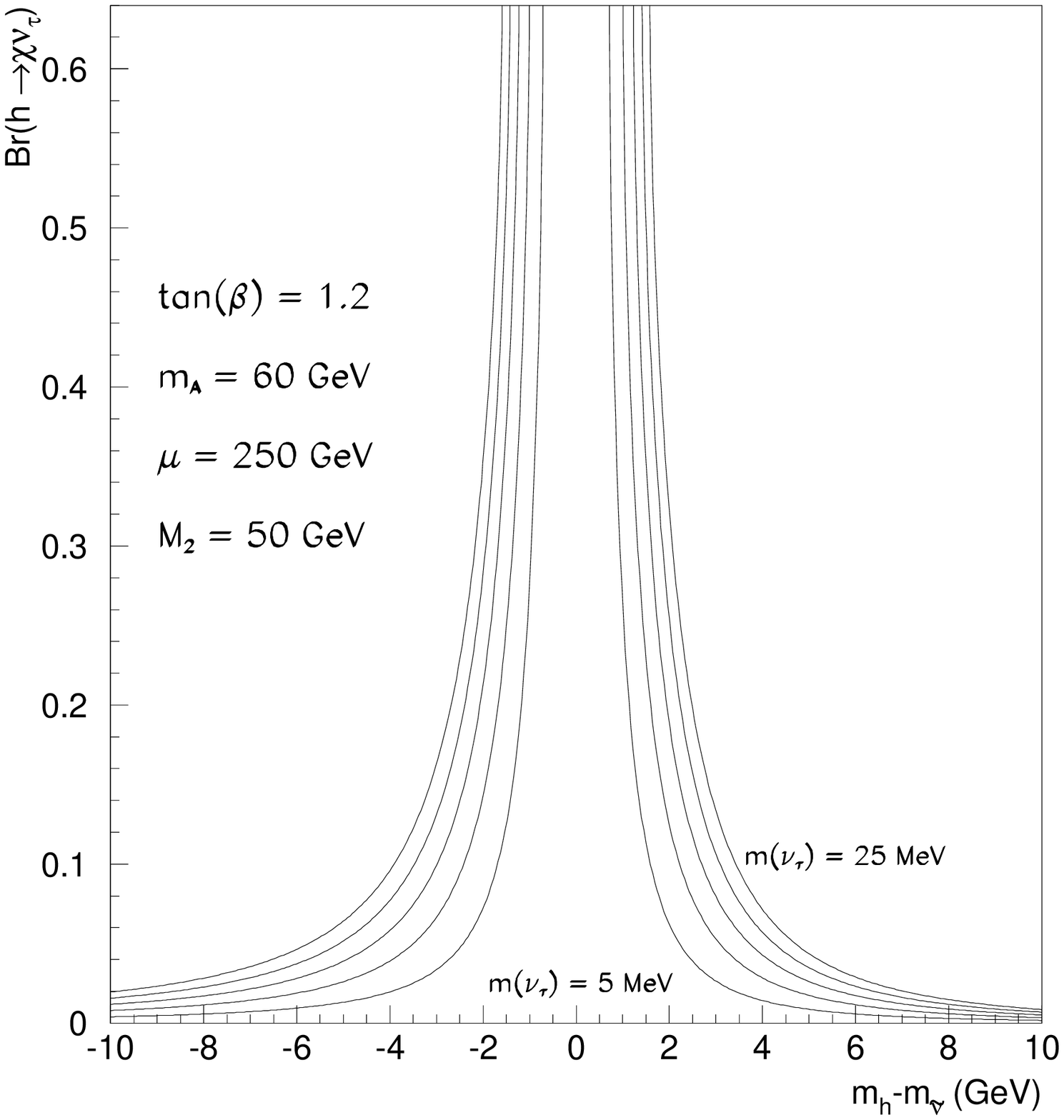,width=\linewidth,height=\linewidth}}
\end{center}  &
\begin{center}
\mbox{
\epsfig{file=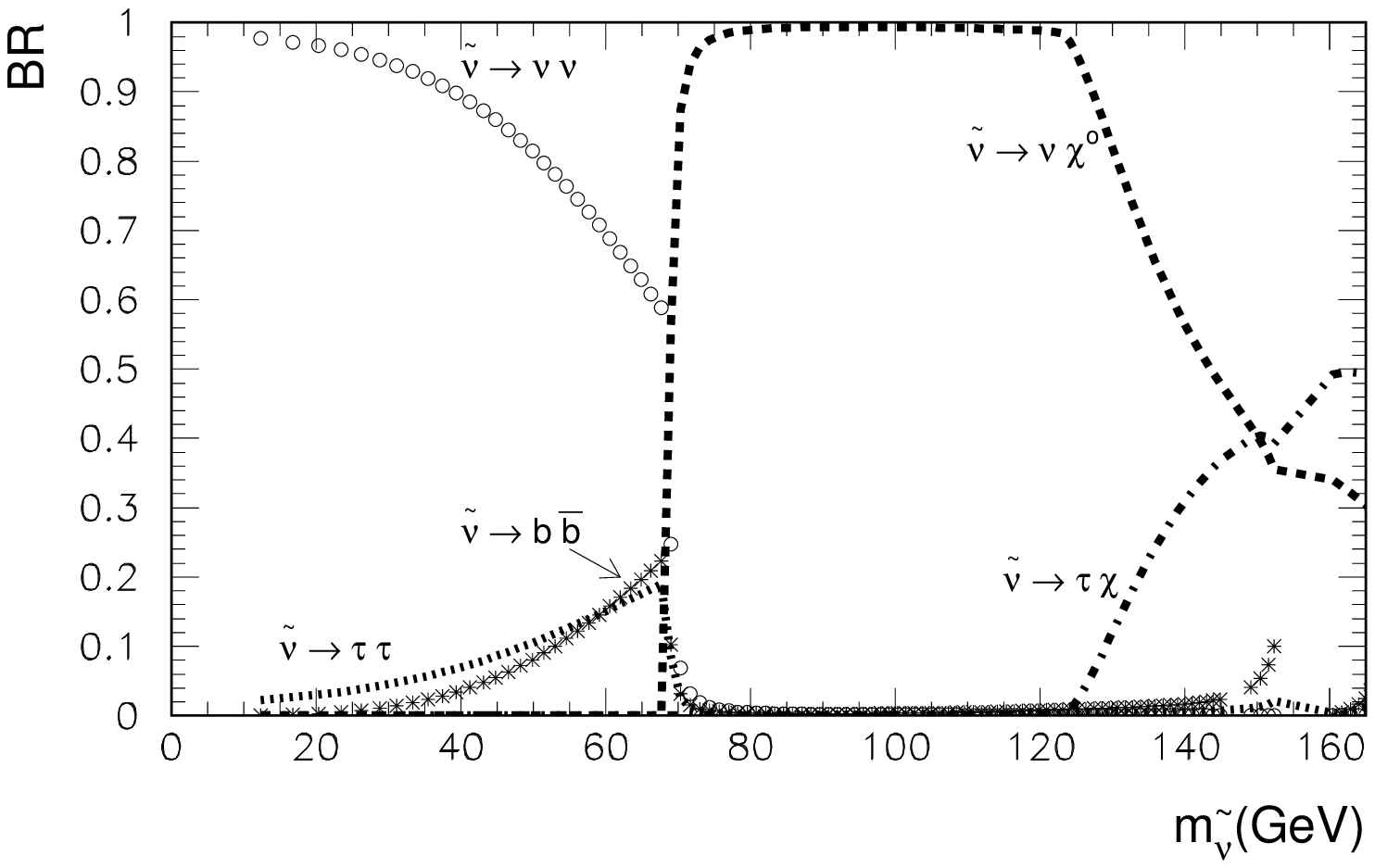,width=\linewidth,height=\linewidth}}
\end{center} \\
\vspace*{-16mm}
\caption{Br$(h \ra \chi^0\nu_{\tau})$ as a function of the difference of
the physical $h$ and $\tilde{\nu}_{\tau}$ masses.
The \nt mass labels the different curves.\label{fig:eps2}} &
\vspace*{-16mm}
\caption{Branching ratios for $\tilde{\nu}_{\tau}$ decays as a function of
its mass for $\tan\beta$=10,$M_2$=70
GeV,$\mu$=200$\,$GeV,$\epsilon_3$=1$\,$GeV,$m_A$=250$\,$GeV.\label{fig:eps5}}\\
\end{tabular}
\vspace*{-12mm}
\end{figure}
\begin{eqnarray}
\label{R}
R_0 = \frac{\Gamma(h\ra \chi^0 \nu_\tau)}{\Gamma(h\ra b \overline{b})}
\approx \frac{\tan^2 \theta_W}{2}\frac{M_W^2}{m_b^2}
\frac{ (1- m_{\chi}^2/M_H^2)^2 }{ (1-4 m_b^2/M_H^2)^{3/2} }
\frac{a_{31}^2}{a_{11}^2} \cos^2\beta \vert\xi\vert^2
\end{eqnarray}
where $\xi$ denotes the appropriate gaugino-LSP mixing element. $R_0$ and
also $R_+ = \frac{\Gamma(h\ra \chi^\pm \tau^\mp)}{\Gamma(h\ra b \overline{b})}$
can be estimated to be of the order of
${\cal O}\left({\epsilon_3^2\over m_b^2}\right)\times{\mu^2\over
(m_h-m_{\tilde\nu_{\tau}})^2}$
In Fig.~\ref{fig:eps2} we display the branching ratios for $h$
decays to LSP + \nt as a function of the $h$-$\tilde{\nu}_{\tau}$
mass difference,
for a suitable choice of SUSY parameters. Clearly, for relatively small
$h$-$\tilde{\nu}_{\tau}$ mass differences of a few GeV, the supersymmetric
channel can dominate over the SM ones (similarly for $A\ra \chi^0\nu_{\tau},
\chi^{\pm}\tau^{\mp}$ decays). Conversely, $\tilde{\nu}_{\tau}$ may decay
into R parity violating SM channels such as
$b \overline{b}$, $\tau^+ \tau^-$ or the invisible mode
$\nu \overline{\nu}$.
These decays are dominant when
the phase space for the R parity conserving channels
such as $\chi \nu$ is closed (see Fig.~\ref{fig:eps5} where
$m_{LSP} \approx 65$ GeV) and may be
non-negligible even above the LSP
threshold if $m_h\approx m_{\tilde{\nu}_{\tau}}$,
leading to a resonant enhancement of
$b \overline{b},\tau^+ \tau^-$ modes
(see small rise of Br($\tilde\nu_\tau  \ra b \overline b$) in
Fig.~\ref{fig:eps5}
for $m_{\tilde{\nu}_{\tau}} \approx m_h = 155$ GeV).
Finally, the R parity breaking terms
lead to the unstable LSP, which would decay inside the detector if
$\epsilon_3 \sim $ few GeV.
Folding the R parity violating Higgs boson decays
to the LSP with the standard decays of $Z^0$
one gets the signatures in $\epm$ collisions which do not occur in the
SM, such as $\tau$$e$ or $\tau$$\mu$ pairs + missing momentum\cite{MY2}.

\section*{Acknowledgements}
\vspace{-2mm}

This work was supported in part by postdoctoral fellowship of the
Spanish MEC and by Alexander von Humboldt Stiftung.


\vspace{-3mm}

\section*{References}
\begin{thebibliography}{99}

\baselineskip 13.9pt

\bibitem{wein}
S.Weinberg, \pr{D26}{82}{533}; N.Sakai, T.Yanagida, \np{B197}{82}{533};
for a review see J.W.F.Valle, \nps {13} {90} {520} and \ib {13} {90} {195}

\bibitem{valleross}
G.G.Ross, J.W.F.~Valle, \pl{B151}{85}{375};
J.Ellis et al, \pl{B150}{85}{142}

\bibitem{mas1}
D.Comelli et al, \pl{B224}{94}{397}

\bibitem{asj2}
A.S.Joshipura, M.Nowakowski, FTUV/94-42, hep-ph 9408224

\bibitem{hall}
L.Hall, M.Suzuki, \np{B231}{84}{419};
I.Lee \np{B246}{84}{120};
K.Enqvist et al, \np{B373}{92}{95}

\bibitem{RPMSW}
J.C.Rom\~ao, J.W.F.Valle \np{B381}{92}{87}

\bibitem{ROMA}
P.Nogueira et al, \pl{B251}{90}{142};
R.Barbieri,L.Hall, \pl{B238}{90}{86};
M.C.Gonzalez-Garcia, J.W.F.Valle, \np{B355}{91}{330};
D.E.Brahm et al, \pr{D42}{92}{1860}

\bibitem{CMP}
Y.Chikashige, R.Mohapatra, R.Peccei, \prl{45}{80}{1926};
for a review see J.W.F.Valle, \ppnp{26}{91}{91}

\bibitem{MY3}
J.C.Rom\~ao, J.Rosiek, J.W.F.Valle, \pl{B351}{95}{497}

\bibitem{MY1}
F.de Campos et al,  \pl{B336}{94}{446}

\bibitem{MY2}
F.de Campos et al, hep-ph 9502237, in print in \np{B}{95}{}

\bibitem{opal}
OPAL collaboration, \zp{C65}{95}{47}

\bibitem{MASI_pot3}
A.Masiero, J.W.F.Valle, \pl {B251}{90}{273};
J.C.Rom\~ao,  C.A.Santos, J.W.F.Valle, \pl{B288}{92}{311}

\bibitem{bj}
J.D.Bjorken, report SLAC-PUB-5673 (1991);
R.E.Schrock, M.Suzuki, \pl{10B}{82}{250};
L.F.Li, Y.Liu, L.Wolfenstein,  \pl{B159}{85}{45};
J.C.Rom\~ao et al, \pl{B292}{92}{329}

\bibitem{alfonso}
A.Lopez-Fernandez et al, \pl{B312}{93}{240};
O.Eboli et al, \np{B421}{94}{65};
D.P.Roy et al, {\sl Phys.Rev.} {\bf D48} (1993) 4224

\bibitem{Aleph92}
ALEPH Collaboration,
{\em Phys.\ Lett.} {\bf B313} (1993) 299; {\bf B313} (1993) 312

\bibitem{aleph1}
ALEPH collaboration, \prep{216}{92}{253}

\bibitem{PDG94}
Particle Data Group, \pr{D50}{94}{1173}

\end{thebibliography}
\end{document}